# The Longest Duration SGRE Event in Solar Cycle 25

Nat Gopalswamy*[1], Pertti Makela[2], Hong Xie[2], Sachiko Akiyama, and Seiji Yashiro[2]
(1) NASA Goddard Space Flight Center, Greenbelt, MD, 20771, USA, https://cdaw.gsfc.nasa.gov
(2) The Catholic University of America, Washington DC 20064, USA

## Abstract

Solar Cycle (SC) 24 was the weakest in the space age, yet it produced many sustained gamma ray emission (SGRE) events from the Sun. Solar cycle (SC) 25, which is a bit stronger than SC 24 observed only a handful of SGRE events over the first five years. Here we report on the 2024 September 14 SGRE event, which has the longest duration (~11.29 hrs) as of this writing. The associated type II radio burst is also of long duration (~16 hr). Detailed analysis of the SGRE event reveals that the event is in good agreement with the linear relation of the SGRE duration with the ending frequency and duration of the type II burst. The kinematics of the associated coronal mass ejection (CME) shows that it is one of the fastest CMEs of SC 25, capable of driving a shock that accelerated >300 MeV protons to account for the observed SGRE. By comparing with an event with similar durations in SC 24, we find that it had a lower-speed CME but resulted in a larger-sized SGRE event. We speculate that the difference may be due to the change in the heliospheric state between the two cycles.

## 1. Introduction

Sustained gamma ray emission (SGRE) from the Sun is characterized by the production of >100 MeV photons well after the impulsive phase where nonthermal particles are accelerated. While the impulsive phase lasts at most for tens of minutes, SGREs last for hours, sometimes for almost a day. Since their discovery in 1985 [1], the SGREs were considered extremely rare in that only a handful of events were reported over the next three decades [2]. After the advent of the Fermi gamma-ray mission, observations from the Large Area Telescope (LAT [3]) have shown that SGREs are rather common [4]. These observations rekindled the debate on the origin of >300 MeV protons precipitating in the photosphere to produce the observed gamma-rays via pion decay. The debate is between the flare and shock origins of the energetic protons. In the former case, energetic protons accelerated during flare impulsive phase remain trapped flare loops [5] sustaining the gamma-ray emission. In the latter, protons accelerated in coronal and interplanetary (IP) shocks propagate back to the Sun to produce the gamma-rays [6]. Energetic coronal mass ejections (CMEs) drive shocks that can remain strong for more than a day accelerating the required energetic protons. The close association of SGREs with energetic CMEs, interplanetary (IP) type II bursts, and solar energetic particle (SEP) events has provided strong support for the shock paradigm [7]. The shock paradigm is further bolstered by backside eruptions that produce gamma-rays on the front side owing to the vast extent of the shock surrounding CMEs [8]. One of the key results in support of the shock paradigm is the linear correlation of the SGRE duration with the ending frequency and the duration of the associated type II bursts [9]. Even though solar gamma-ray activity was high in solar cycle (SC) 24, the number of long-duration SGREs is relatively small in SC 25 over the first 5 years. Given the significant difference between SCs it is important to check whether the SGRE – type II relation holds good in SC 25.

In this paper we report on the 2024 September 14 SGRE event, which turns out to be the longest event so far in SC 25. We also check other characteristics such as an ultrafast CME and long-lived type II radio burst. We compare our event with another event of similar durations in SC 24 to get a clue on the solar cycle variation of SGRE events.

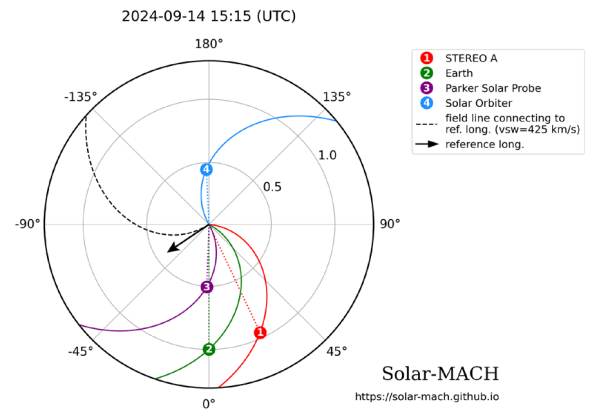

**Figure 1.** Relative locations of various spacecraft that observed the 2024 September 14 eruptive event whose longitude was E56 (indicated by the black arrow). STA was at W25 and the eruption was a limb event in STA view. Parker Solar Probe (PSP) was very close to the Sun-Earth line (W4) located at a distance of 109 Rs.

## 2. Observations

The 2024 September 14 eruption was well observed by multiple spacecraft from near the Sun-Earth line. The Solar and Heliospheric Observatory (SOHO), the Solar Dynamics Observatory (SDO), Wind, ACE, and DISCOVR were all located along the Sun-Earth line. The

Solar Terrestrial Relations Observatory (STEREO)-Ahead (STA) was located at W25. The eruption occurred in NOAA active region (AR) 13825 located at S15E56. In STA field of view (FOV), the location is S15E81, very close to the limb so the height measurements in STEREO coronagraphic FOV have minimal projection effects.

**2.1 Solar Source and CME Kinematics**

Figure 2 shows the X4.5 flare from GOES and an SDO image of the flare/heated prominence. The SDO image was obtained by the Atmospheric Imaging Assembly (AIA [10]) at 94 Å. The emission comes from a 6 MK plasma in the flare structure. The eruptive prominence ejected from reconnection region is heated at least to 6 MK because of the emission at 94 Å. The CME first appeared above the east limb in STA COR1 FOV [11] at 15:21 UT with the leading edge at a height of 1.78 Rs. In the next two images, the LE moved to 2.57 Rs (15:26 UT) and 3.34 Rs (15:31 UT). The first appearance was later at 15:36 UT (4.81 Rs) in the Large Angle and Spectrometric Coronagraph (LASCO [12]) FOV indicating a rapid acceleration of the CME. The CME already had a width of 95º at first appearance and became a full halo at 17:06 UT.

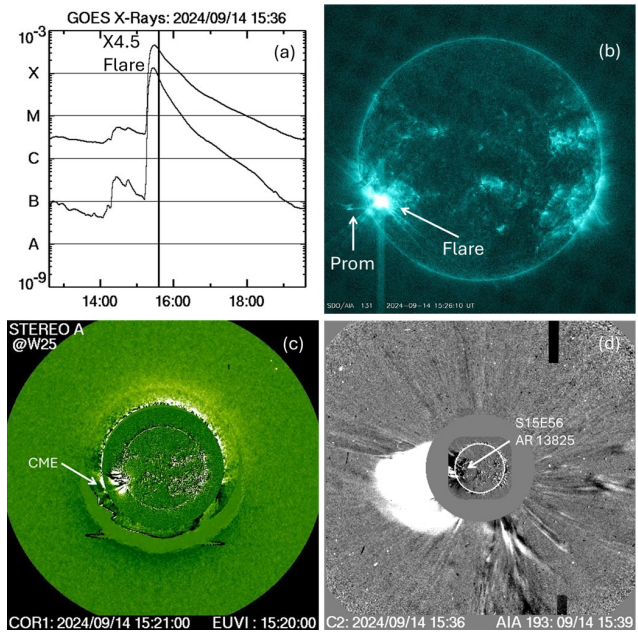

**Figure 2.** (a) Light curve of the GOES X4.5 flare that started, peaked and ended at 15:13, 15:29, and 15:47 UT, respectively. (b) EUV flare observed by SDO/AIA at 94 Å, which corresponds to 6 MK plasma. The flare was saturated in the SDO image, and an eruptive prominence can be seen above the eruption site in the 94 Å emission. (c) The CME first appeared in the STA/COR1 FOV with a superposed EUVI image. The CME leading edge is at a height of 2.57 Rs. (d) The first appearance of the CME in LASCO/C2 FOV. The LE is already at a height of 4.81 Rs.

Figure 3 shows the height-time history of the CME, whose nose is approximately at position angle (PA) = 95º. A linear fit to the height-time data points gives an average speed of 2366 km s$^{-1}$. Since the source location is at S15 E56, the deprojected speed is ~2534 km s$^{-1}$, making it one of the fastest CMEs of SC 25. A second order fit gives an average positive acceleration of ~0.11 km s$^{-2}$, which is the residual acceleration. The average initial acceleration can be derived from the soft X-ray flare duration (16 min) and the average speed (2366 km s$^{-1}$) as in the coronagraph FOV as 4.9 km s$^{-2}$ and the deprojected value is 5.95 km s$^{-2}$. Such high initial acceleration is an indication that the CME should be accelerating protons to very high energies with a hard spectrum [13]. The average acceleration within the coronagraph FOV is ~0.1 km s$^{-2}$, which does not include the initial acceleration. By the time the CME reached ~20 Rs in the C3 FOV, the speed increased to 2713 km s$^{-1}$. Thus, the 2024 September 14 CME belongs to the class of energetic CMEs that are known to produce ground level enhancement (GLE) in SEP events and SGRE events.

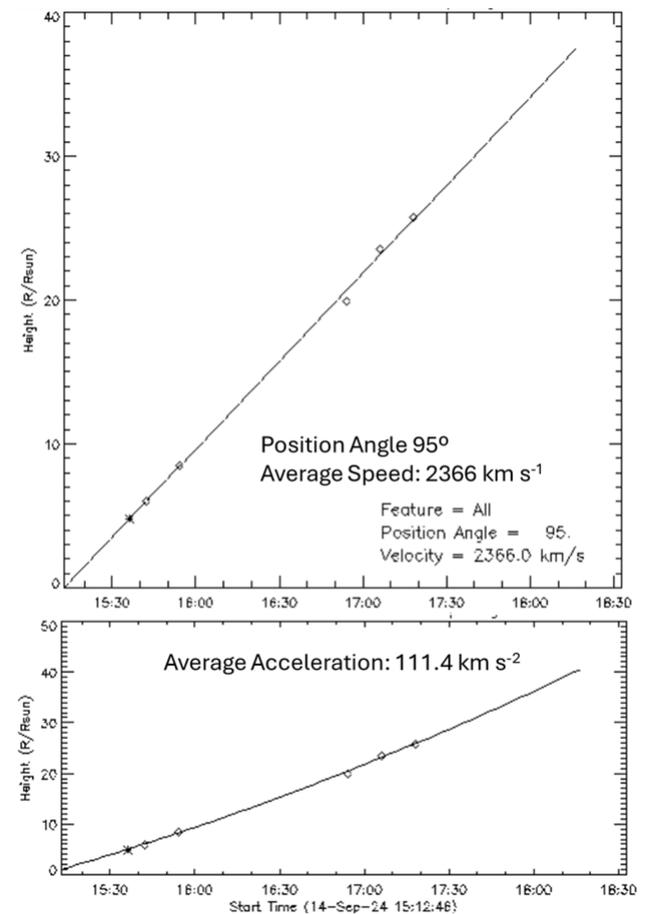

**Figure 3.** Linear (top) and second order fit (bottom) to the height-time measurements from LASCO C2 (asterisk) and C3 (diamonds). The linear fit gives an average speed within the coronagraph FOV as 2366 km s$^{-1}$. The average acceleration within the coronagraph FOV is positive, indicating continued CME acceleration in the FOV.

**2.2 THE SGRE and Type II Radio Burst**

The Sun was in the Fermi/LAT FOV during the 2024 September 14 eruption. Figure 4 shows the time evolution of the >100 MeV gamma-ray flux. The SGRE peak is around the peak of the GOES soft X-ray (SXR) flux. There was no data during the rise phase of the SGRE because LAT does not observe the Sun continuously. Our definition of the SGRE duration is the interval from the peak of the associated soft X-ray flare to the midpoint between the last SGRE point above the background and the next data point [9]. Therefore, the lack of observation for a few minutes before the SGRE peak does not affect the SGRE duration. The SGRE duration from Fig. 4 is $11.29 \pm 1.57$ hrs. The peak gamma-ray flux is $1.95 \times 10^{-4}$ photons cm$^{-2}$ s$^{-1}$. This is one of the highest flux events of SC 25. The 2024 February 9 event had the highest flux of $3.91 \times 10^{-4}$ photons cm$^{-2}$ s$^{-1}$, which is only higher by factor of 2. The gamma-ray fluence is 0.89 cm$^{-2}$, which is above the average of cycle-24 SGRE fluences.

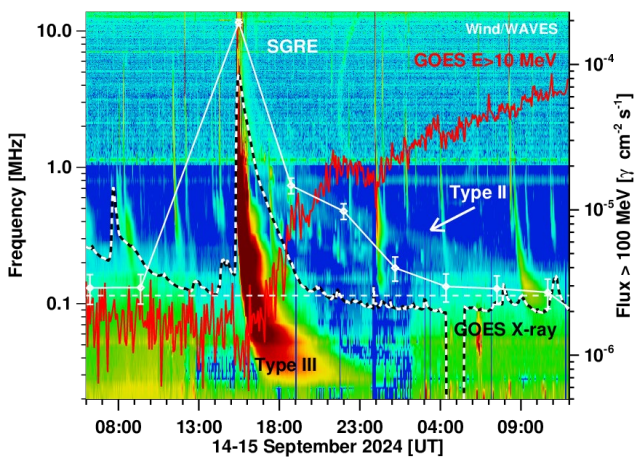

**Figure 4.** Wind/WAVES dynamic spectrum showing type III and type II bursts with superposed light curve of >100 MeV gamma-ray flux (white curve, marked SGRE), GOES soft X-ray intensity (dashed black and white curve), and the >10 MeV proton flux from GOES (red curve). Blue is the weakest intensity and red is the highest. The fuzzy feature between the SXR and SGRE curves is the beginning of the IP type II burst that lasts for several hours until its end just before 9 UT on September 15. The type III burst starts slightly after the onset of the GOES flare.

In the Wind/WAVES [14] radio dynamic spectrum in Fig. 4 the eruption-related type III bursts are the most intense bursts. Type III activity began slightly after the SXR flare onset and ended around the SXR peak, implying that the bursts are due to electrons accelerated in the impulsive phase of the flare that escape along open field lines into the IP medium. The type II burst is relatively weak, but had fundamental, second harmonic, and third harmonic structure, which is rare at these frequencies [15]. One can discern the fundamental and second harmonic in Fig. 4 in the radio features between the GOES SXR and SGRE curves. The type II burst continues into the next day, ending somewhere between 6:30 and 8:30 UT on September 15. Noting that the burst started around 15:30 UT, we get a duration of $16 \pm 1$ hrs. By the time the type II burst ended, it drifted down to 170 kHz. At the higher frequency side, the eruption had metric type II bursts. Thus, this burst has emission components from metric (m) to kilometric (km) wavelengths. Such m-km type II bursts are characteristic of energetic eruptions resulting in GLEs [15] and SGRE [9] events.

## 3. Analysis and Results

From SC 24 SGRE observations it was found that SGRE durations are inversely proportional to the ending frequency of type II burst and directly proportional to the type II burst duration [9]. Figure 5 shows these relations in the form of scatter plots. It is remarkable that the gamma-ray emission ends when the type II burst ends despite the fact that the former is due to >300 MeV protons, while the latter are due to lower energy electrons but accelerated by the same shock. The data points of the 2024 September 14 events, viz., (11.29, 170) and (1.29, 16.0) are shown by the red dots in Fig. 5a and 5b, respectively. We see that the durations are in remarkable agreement with the relations obtained in SC 24. Furthermore, this event fills a gap in the scatter plot and makes the relation even more robust.

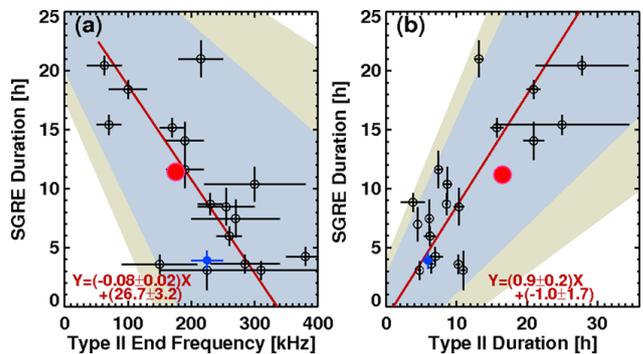

**Figure 5.** (a) Scatter plot between SGRE duration and ending frequency of type II bursts with the 2024 September 14 event shown as a red data point. (b) Scatter plot between SGRE duration and type II burst duration with the 2024 September 14 event shown as a red data point. The blue and yellow shaded areas denote the 95% and 99% confidence levels. The linear relations were derived using 19 SGRE events that had durations >3 hrs in SC 24 [9].

We also compared the flare and CME properties with the statistical relation obtained before [9]. The SGRE fluence Fs and CME speed (V) are related by Fs = V$^6$/100, where V is in units of 1000 km s$^{-1}$. For our event, V = 2.534, we get Fs = 2.65 cm$^{-2}$, which is a factor of 3 higher than the observed value, but within the scatter of the data points. Fs and SXR fluence (Fx) are related by: Fs = 6.31Fx$^{1.6}$. NOAA/SWPC event report lists Fx = 0.58 J m$^{-2}$, so we get Fs = 2.64 cm$^{-2}$, again higher than the observed value. Finally, Fs and the SXR peak intensity Wx are related by: Fs = $2.0 \times 10^5$ Wx$^{1.4}$. For Wx = $4.5 \times 10^{-4}$ W m$^{-2}$ gives Fs = 4.13. In all the three cases, the fluences seems to be less than what is expected from the empirical relations derived from SC 24 SGREs, although the data points are well within the scatter of the parameters.

## 5. Discussion and Summary

We reported on the longest-duration SGRE event of SC 25 as of this writing. The SGRE event was relatively strong with a peak flux of $\sim 1.95 \times 10^{-4}$ photons cm$^{-2}$ s$^{-1}$. The peak flux is only a factor of 2 smaller than the highest peak flux in SC 25. The gamma-ray fluence is $\sim 0.89$ cm$^{-2}$, which is an above-average value when compared to all the Fermi/LAT SGRE fluences in SC 24 [9]. The CME was ultrafast with a sky-plane speed of 2366 km s$^{-1}$ with the deprojected speed of $\sim 2534$ km s$^{-1}$. For comparison, the average deprojected speed of 20 SGREs from SC 24 is 2000 km s$^{-1}$. The CME was a full halo even though it is close to the limb; limb halos are quite energetic.

It is worth comparing our event with the 2014 Feb 25 event in SC 24 that had similar SGRE duration, peak flux, and fluence [9]. The SGRE lasted for 8.46 hrs, while the type II burst lasted for 10.32 hrs. The CME was an ultrafast (2153 km s$^{-1}$) full halo accompanied by an X4.9 flare originating from AR 11990 located at S12E82. The peak flux and fluence were $1.62 \times 10^{-3}$ photons cm$^{-2}$ s$^{-1}$ and 14.9 cm$^{-2}$, respectively, each an order of magnitude higher than the corresponding values of our event. These values are even higher by a factor of 4 when corrected for the limb location (E82) because part of the extended gamma-ray source may be behind the limb [15]. The flare sizes are similar. Thus, the SC 24 SGRE is larger than ours even though our CME is faster (2534 km s$^{-1}$ vs. 2153 km s$^{-1}$).

A preliminary comparison between the first five years of observations in SCs 24 and 25 in terms of the number of SGRE events, duration, and size yields some puzzling results. The number of SGRE events with duration >3 hr dropped by $\sim 57\%$ (6 vs. 14). The average duration of SC 25 events also dropped from $\sim 6.3$ in SC 25 compared to 12.1 hr in SC 24 (a 48% reduction). This is in contrast to the fact that SC 25 is slightly stronger than SC 24. While the SC 25 events agree with the relation between SGRE and type II bursts, they are clustered toward the lower end of the axes. One possibility is the change in Alfven speed in the corona and IP medium that decides the shock strength for a given CME speed. It is known that the ambient Alfven speed can vary by a factor of $\sim 3$ [16]. There were some changes in the coverage of the Sun due to instrument issues, which also need to be considered in understanding the discrepancy between the two SCs.

In summary, the relation of SGRE duration with the ending frequency and duration of the associated II bursts derived from SGRE events of SC 24 holds good for the 2024 September 14 SGRE. A comparison of our event with the one on 2014 February 25 suggests that the 2024 event had a lot more energy but produced only a moderate-size SGRE event. Further work is needed to see if the difference between the two cycles can be attributed to the changing Alfven speed distribution in the corona and IP medium.

## 7. Acknowledgements

This study benefited from the open data policy of NASA and NOAA. NG is supported by NASA's STEREO project and the Living With Star program. PM is supported by NSF grant AGS-2043131 HX is supported by NSF grant AGS-2228967.